\documentstyle[12pt,aaspp4]{article}

\begin{document}

\title{Quasi-regular X-ray bursts from GRS 1915+105 observed with the IXAE:
possible evidence for matter disappearing into the event horizon
of the black hole }

\author{B. Paul, P. C. Agrawal, A. R. Rao, M. N. Vahia, and J. S. Yadav}
\affil{Tata Institute of Fundamental Research, Homi Bhabha Road, Mumbai 400 005, India}
\affil{e-mail: bpaul@tifrvax.tifr.res.in (BP), pagrawal@tifrvax.tifr.res.in (PCA),
arrao@tifrvax.tifr.res.in (ARR), vahia@tifrvax.tifr.res.in (MNV) and jsyadav@tifrvax.tifr.res.in (JSY)}
	\and
\author{S. Seetha and K. Kasturirangan}
\affil{ISRO Satellite Centre, Airport Road, Vimanpura P.O. Bangalore  560 017, India.}
\affil{e-mail: seetha@isac.ernet.in (SS)}

\begin{abstract}

Three different types of very intense, quasi-regular X-ray bursts have
been observed from the Galactic superluminal X-ray transient source
GRS~1915+105 with the Pointed Proportional Counters of the
Indian X-ray Astronomy Experiment onboard the Indian satellite IRS-P3.
The observations were carried out from 1997 June 12 to June 29 in
the energy range of 2$-$18 keV and revealed the presence of persistent
quasi-regular bursts with different structures. Only one of the three
types of bursts is regular in occurrence revealing a stable profile
over extended durations. The regular bursts have an exponential rise
with a time scale of about 7 to 10 s and a sharp linear decay in 2 to 3 s.
The X-ray spectrum becomes progressively harder as the burst evolves
and it is the hardest near the end of the burst decay. The profile
and energetics of the bursts in this black hole candidate source are distinct
from both the type I and type II X-ray bursts observed in neutron star
sources. We propose that the sharp decay in the observed burst pattern
is a signature of the disappearance of matter through the black hole
horizon. The regular pattern of the bursts can be produced by material
influx into the inner disk due to oscillations in a shock front far
away from the compact object.

\end{abstract}

\keywords{accretion, accretion disks --- binaries: close ---
black hole physics --- X-rays: bursts, stars --- stars: individual GRS~1915+105}

\section{Introduction}

The X-ray transient source GRS~1915+105 was discovered in 1992 with
the WATCH all sky X-ray monitor onboard the GRANAT satellite
(\cite{cast:94}). During the two years of hard X-ray observations by
WATCH, two powerful outbursts were discovered and during the peak of
the outbursts the source luminosity was as high as $10^{39}$ erg s$^{-1}$.
Superluminal motions of two symmetric radio emitting jets
of GRS~1915+105 were discovered by Mirabel \& Rodriguez
(\cite{mira:94}). Correlated enhanced radio and X-ray emissions were
discovered from the source with a near simultaneous monitoring over a
long period (\cite{fost:96}; \cite{harm:97}).

The source went into a very high X-ray luminosity state in early 1996
and was observed on several occasions by the Pointed Proportional
Counters (PPCs) of the Indian X-ray Astronomy Experiment (IXAE)
(\cite{agra:97}; \cite{paul:97a}), the Proportional Counter Array (PCA) and the All Sky
Monitor (ASM) of the Rossi X-ray Timing Explorer (RXTE) (\cite{brad:96}).
The X-ray intensity was found to vary on a variety of time scales and
the spectrum also changed during the brightness variations (\cite{grei:96}).
PPC observations of GRS~1915+105 in its low hard state in 1996 July
showed intensity variations by a factor of 2 to 3 at 100$-$400 ms time
scale (\cite{paul:97a};b). Strong (rms variability 9\%) and narrow
(${\nu \over \delta \nu} \approx 5$) Quasi Periodic Oscillations (QPOs) of
varying frequency were discovered in GRS~1915+105 with the PPC
observations (\cite{agra:96}). Intensity dependent narrow QPOs were also
detected with the PCA (\cite{morg:97}).

Several features in the observed properties of GRS~1915+105 such as the
Power Density Spectra (PDS) with the QPO feature, a hard X-ray tail and
the subsecond time variability, are typical characteristics of black hole
binaries. The X-ray intensity is found to be more than
$10^{39}$ erg s$^{-1}$ for extended periods which is super-Eddington
luminosity for a neutron star. Stable and narrow QPOs at a frequency of
67 Hz were observed with the PCA, and associating these to the Keplerian
period of the innermost stable orbit around a black hole, the mass for a
non-rotating black hole has been estimated to be 33 M$_\odot$ (\cite{morg:97}).

The most compelling evidence for the existence of a black hole in Galactic
X-ray binaries normally comes from the measured mass function which indicates
that the mass of the compact object is much
larger than that permitted for a neutron star. In the absence of measured
binary parameters (like in the case of GRS~1915+105) phenomenological 
arguments are normally used, which, though compelling for a class of 
objects, are not conclusive enough for individual cases. This is mainly due
to the fact that the accretion disk around a black hole has properties
quite similar to that around a low magnetic field neutron star (\cite{tana:95}).
Recent progress in the understanding of accretion onto black holes,
however, has indicated a possible way of uniquely separating the properties
of accretion onto black holes and neutron stars. It is found that the black
hole accretion disks are cooled by advection in their innermost parts and 
it has been realized that advection is one of the most fundamental features of 
the black hole accretion (\cite{chak:96}; \cite{abra:97}).  
In this {\it Letter} we present a possible evidence for the direct
detection of advection in GRS~1915+105. This is based on the detection of
regular and persistent X-ray bursts from this source. The bursts are
different in temporal structure and regularity of occurrence from the
classical bursts in Low Mass X-ray Binaries (LMXB). The bursts have a slow exponential rise
and sharp decay. The sharp decay, which is the most significant
difference from the bursts in the neutron star sources, and the hardening
of the spectrum as the burst progresses, indicate a possibility that we
are observing the advection of matter into the black hole horizon. In
the following sections, we describe the observations and properties of
635 regular bursts observed with the PPCs. We discuss the possibility
of such a phenomena taking place in the framework of an accretion disk
model with an oscillating shock front (\cite{chak:95}). 

\section{Observations}

The observations were carried out using the 3 PPCs of the Indian X-ray
Astronomy Experiment (IXAE) onboard the Indian satellite IRS-P3 launched
on 1996 March 21 from India using a PSLV rocket. The principle
objective of this experiment is to carry  out short and long term
variability studies of X-ray binaries and other variable X-ray sources.
The PPCs, filled with argon-methane mixture at 800 torr pressure and
working in the 2$-$18 keV energy range, have a total area of
1200 cm$^2$ and field of view of $2.3^\circ \times 2.3^\circ$. The
energy resolution is $\approx 22{({E\over6})}^{-{1\over2}}\%$ at E keV
with a detection efficiency of about 65\% at 6 keV and 10\% at 15 keV.
Each PPC is a multilayer unit consisting of 54 anode cells of size
1.1 cm $\times$ 1.1 cm arranged in 3 identical layers.
The end cells  of  each 
layer and all the 18 anodes of the third layer are connected  together 
and  operated as a veto layer for the top two layers which  constitute 
the X-ray detection volume. The alternate anodes in each of the two
X-ray detection layers are joined together and operated in mutual
anti-coincidence to reject charged particle induced background.
A star tracker onboard the IRS-P3 satellite
co-aligned with the viewing axes of the proportional counters is used
for pointing towards the X-ray sources. For further details of the
PPCs and the observation methodology see Rao et al. (\cite{rao:97}).

Observations with the PPCs are usually made in about 5 orbits of
the satellite every day and each observation has a duration of about
20 minutes. During the observation period of 1997 June 12 to June 29,
a total of 39,300 seconds of useful exposure time was obtained on source.
Data were recorded with a time resolution of 1 s during June 21$-$26 and
0.1 s on the rest of the days. About half of the useful exposure was
obtained with 1.0 s time resolution. In the 1.0 s mode of observation,
data are available for five orbits every day whereas in the 0.1 s mode
usually data are available only for three orbits due to the limited
size of the onboard data storage unit.

\section{Results}

Three types of bursts are observed during the PPC observations over the
period of 1997 June 12 to 29 - (a) regular bursts, having a slow rise
and sharp decay lasting for $\sim$ 10 s and recurring every 45 s,
(b) irregular bursts of variable duration, slow rise, flat top and sharp
decay, and (c) long bursts, with duration of a few tens to a few hundred
seconds, followed by sharp decay. Sharp decay is a common feature of all
the bursts. Regular bursts were detected during June 12$-$17 and
again during June 22$-$26, the irregular bursts during June 18$-$21
and long bursts were detected after June 27. Representative light curves
of 500 s duration
obtained on different days are shown in Figure 1. All the panels in the
figure have similar Y-axis scales, and the X-axis is adjusted
such that a burst decay phase occurs around 250 s. A secondary peak
near the end of the bursts is a common feature of all the bursts. A total 
of 635 regular bursts (in $\sim$ 28,200 s of observation), 78 irregular
bursts (in 6,200 s) and 40 long bursts (in $\sim$ 4,900 s) have been detected.
The long bursts show higher variability near the end of the burst and
the burst duration is correlated to the quiescent state period just
prior to the burst. Similar behavior is also reported from PCA observations
carried out in 1996 June (\cite{bell:97a}). Several irregular bursts,
concurrent with the present observations on 1997 June 18 and having 
similar properties has also been detected in the PCA data (\cite{bell:97b}).
\cite{taam:97} reported the detection of a series of regular bursts recurring
every 60 s to 100 s during 5.5 hours of PCA observation on 1996 October 15.
The present observations show that the regular bursts are stable for several
days. We discuss the properties of these bursts below.

The regular bursts detected during June 12$-$17 and again during
June 22$-$26, have a peak intensity of about 3 to 5 times the quiescent
intensity. In all the bursts, a dip is present just before the decay of
the burst. But the most remarkable feature of our observations is the
persistence of the regular bursts for a few days with similar shape,
structure and period.
The separation between the successive bursts shows a random walk in time
instead of any regular pattern. Time separation between the bursts
averaged over one day is found to be in the range of 40 s to 52 s, with a
large scatter. The distribution of burst interval for each day fits well
with a Gaussian, with a tail on the higher side, having a mean in the
range of 40 s to 50 s and $\sigma \sim$ 3 s.

Individual bursts are well fitted with a profile
which is a sum of two bursts with exponential rise and very fast
linear decay along with a constant emission. A typical burst
profile is shown in Figure 2.
To improve the statistical accuracy of the data we have co-added
a large number of bursts by matching the peak of the fitted
profiles. The co-added burst profiles in two different energy
ranges (2$-$6 keV and 6$-$18 keV) are shown in the top two panels
while the hardness ratio is shown in the third panel of Figure 3.
The sharp features are smeared
due to addition of bursts of different duration. Intensity changes
are more prominent at higher energy and the energy spectrum becomes
harder as the burst progresses. The burst is hardest near the end
of its decay. This is a unique feature of these bursts which
distinguishes them from the bursts seen in LMXBs which become
softer in the decaying phase (\cite{lewi:95}).

We have calculated the possible temperature and radius distribution
assuming a multi-temperature disk black-body model for the burst
emission with the temperature (kT$_{in}$) and radius (R$_{in}$)
of the innermost region of the accretion disk as free parameters.
For this purpose, the burst count rate and the hardness ratio are
obtained after subtracting the quiescent value from the observed 
count rates. From the spectral fitting reported by \cite{taam:97},
it is seen that during a burst the power-law component in the
2$-$10 keV region changes by a factor less than 2, whereas the disk
emission component changes by a factor more than 4. This indicates that
the burst emission has a spectral type which is similar to the disk
black-body emission. From the response matrix of the PPC detectors, the
observed total count rate and hardness ratio profiles are converted to the
R$_{in}$ and kT$_{in}$ of the innermost disk.
The hardness ratio profile of only the burst component,
temperature of the inner part of the disk (kT$_{in}$) and the inner radius
of the disk (R$_{in}$) are shown in the three panels at the bottom
of Figure 3. The data after 30 s are taken as the quiescent level and
hence they are not used for the hardness ratio calculation and plotting.
It can be seen from the figure
that the temperature increases sharply during the burst decay phase. The
radius R$_{in}$ remains constant during most of the rising phase of the
burst, but decreases sharply during the burst decay.

\section{Discussion}

The bursts in GRS~1915+105 are very different from the type I
X-ray bursts seen in about 40 LMXBs and type II X-ray bursts 
in the Rapid Burster (MXB 1730$-$335) because of their unique
feature of slow rise and fast decay. All the bursts in the LMXBs have
fast rise time of less than a second to 10 seconds and slow
decay of 10 seconds to a few minutes (\cite{lewi:95}).
The type I X-ray bursts are understood to be
thermo-nuclear flashes caused by accretion of matter on to the
neutron star surface. The type II bursts are caused by sudden infall of
matter on to the neutron star due to some instability in the
inner part of the disk supported by the magnetic field.
The slow decay of the
burst intensity represents the cooling time scale of the neutron
star photosphere.

The two peculiarities of the bursts in GRS~1915+105 are the
regularity with which they occur over time scale of several days, and
the presence of a secondary peak in all the bursts. In the classical
bursts, the spectrum is initially hard and becomes softer as
the burst decays (\cite{lewi:95}). In sharp contrast, the
bursts in GRS~1915+105 remain hard till the end
and it is, in fact, the hardest near the end of the burst.

The ratio of luminosity in type I X-ray bursts ($L_b$) and
the average quiescent X-ray luminosity ($L_p$) is
$ {L_b\over L_p} \sim 10^{-2}$. The time-averaged
type II burst luminosity is much higher, usually 0.4 to 2.2 times the
average luminosity of quiescent emission (\cite{lewi:95}).
The time-averaged luminosity of the regular bursts detected from
GRS~1915+105 is 0.3 to 0.5 times the luminosity of the quiescent
emission. This is much higher than the ratio in type I bursts
(where the thermonuclear process has much smaller efficiency compared
to the gravitational process) and less than the type II bursts (where
the burst emission is due to gravitational energy release).
The emission process involved in producing the bursts here is not
likely to be thermo-nuclear because of the energetics involved.
If the energy generation process is gravitational (like in type II
bursts), the difference in efficiency might indicate the absence of
hard surface in the compact object.
A process in which the energy produced is due to
gravitational potential but not all
the energy is emitted as radiation, part of it being advected into the
event horizon as kinetic energy of the matter, is appropriate
for this source.

\cite{taam:97} have attempted to describe the regular bursts in
GRS~1915+105 in the frame work of the thermal/viscous instabilities
in the accretion disks. This model is based on the fact that the inner
regions of accretion disk in the standard model are unstable to thermal
and surface density fluctuations (\cite{taam:84}). Using various 
scaling laws for viscosity (like the viscous stress scaling as the
total pressure), they solved time dependent differential
equations for the inner accretion disk and found that the thermal
instability manifests as short duration luminosity fluctuations or bursts.
The accretion disk models which 
solve the inner boundary conditions and invoke the advection effects
explicitly (\cite{chak:96}; \cite{abra:97}), however, find that the
thermal/viscous instabilities are removed completely by the addition
of advection effects.
Hence we attempt to explain the regular bursts in GRS~1915+105 in
terms of advection effects in an accretion disk around a black hole.

The regular bursts observed in GRS~1915+105 can be due to periodic
infall of matter onto a black hole from an oscillating shock front.
In the black hole accretion disk model of Chakrabarti and Titarchuk (1995),
the disk
has two components, an equatorial Keplerian disk and a sub-Keplerian
component just above and below the Keplerian disk. The sub-Keplerian
component of the disk experiences a shock due to centrifugal barrier
and if the cooling time scale of the post-shock halo matches with the
material infall time scale, oscillations can set in (\cite{molt:96}).
If the matter accreted from the companion has high angular momentum
and low viscosity, the shock front can be far away from the black hole
and oscillation period can be very large. The oscillation period ($t_{osc}$)
of the shock depends on the mass of the black hole ($M$) and the radius at
which the shock is formed ($R$),  and is given by
$t_{osc} = 12.5~{{({R\over{1000~r_g}})}^{3\over2}}~({M\over{10~M_\odot}})$~s
where $r_g$ is the Schwartzchild radius (\cite{chak:97}).
For a 30 M$_\odot$ black hole, if the oscillation period
is 50 s, the shock front should be at a radius of 1200 $r_g$.

It is possible that at some particular phase of this oscillation the
piled up matter behind the shock falls catastrophically onto the
black hole and a burst is produced. As the matter goes in, temperature
increases producing large X-ray intensity. The free fall time scale, from
the shock front at a distance R, to the black hole is
$t_{free} = 2.1~{{({R\over{1000~r_g}})}^{3\over2}}~({M\over{10~M_\odot}})$~s.
Comparing $t_{osc}$ and $t_{free}$ we find that for $t_{osc}$ to be about
50 s, the free fall time should be 8.4 s in fair agreement with the rise
time of the bursts. The burst is suddenly terminated as the matter goes
behind the event horizon of the black hole. In this scenario, as the burst
progresses, the temperature of the infalling matter increases, giving
rise to the observed spectral hardening. In the same model, the irregular
and long bursts can be produced, if the accreted matter has somewhat
larger angular momentum, so that a momentary disk is formed before it
is advected.

\section{Conclusion}

In conclusion, we have presented observations of a unique
type of X-ray bursts in the galactic superluminal transient source
GRS~1915+105. This is the only black hole candidate in which
regular bursts are observed. The observed bursts are very different 
compared to the classical bursts in the LMXBs both in terms of temporal
structure and spectral evolution. We propose that the sharp decay
of the bursts and the hardening of the spectrum near the end of the
decay indicate disappearance of accreted matter into the black hole
horizon.

Since the advection is one of the fundamental features of
accretion onto black holes, such temperature profiles during outbursts
in black hole candidates should be commonly observable and provide an
observational test to distinguish the black hole binaries from the neutron
star binaries. For example, the co-added shots in the well known
Galactic black hole candidate Cygnus X-1 shows spectral hardening after
the burst peak (\cite{nego:94}), which possibly indicates advection effect. A
quantitative fitting of profile of similar bursts using the refined
accretion disk models and investigating similar effects in other black
hole candidate sources should eventually provide an unique
distinguishing feature of accretion onto the black holes.

\begin{acknowledgements}

We wish to thank an anonymous referee for thoughtful suggestions which
considerably improved the paper. We also thank K. P. Singh for the valuable
comments on the manuscript and S. K. Chakrabarti for information about
the multi component accretion disk model. We acknowledge K. Thyagrajan,
Project Director of IRS-P3, R. N. Tyagi, Manager PMO and 
R. Aravamudan, Director, ISAC for their support. The valuable
contributions of the technical and engineering staff of ISAC and TIFR in
making the IXAE payload are gratefully acknowledged.

\end{acknowledgements}

\newpage

\begin{figure}
\vskip 15cm
\includegraphics{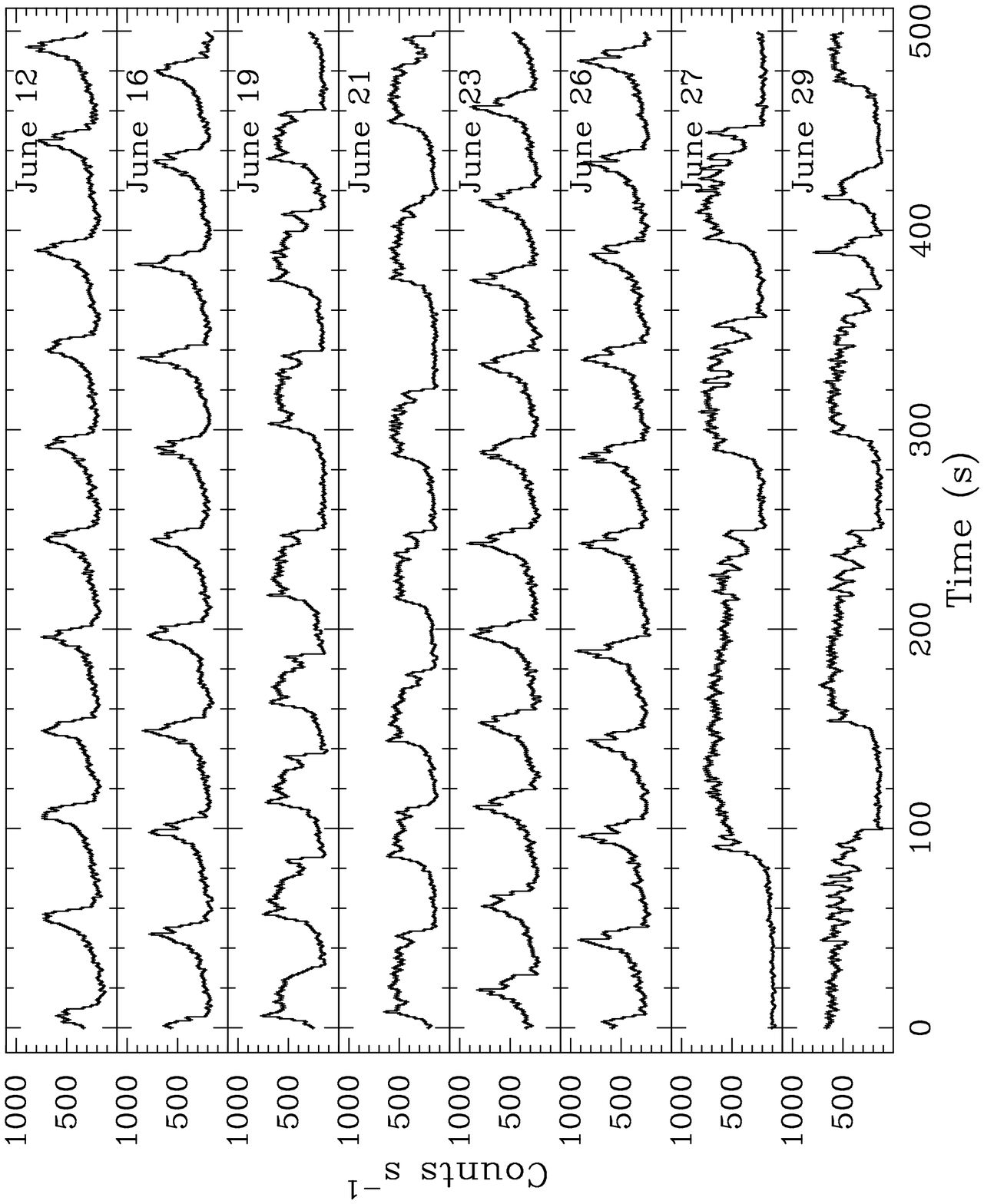} 
\caption{The regular (first, second, fifth and
sixth panel from the top), irregular (third and fourth) and
long (seventh and eighth panel) bursts observed in GRS~1915+105 with
one of the PPCs. Date of each observation is given in the respective
panels.}
\end{figure}

\begin{figure}
\vskip 15cm
\includegraphics{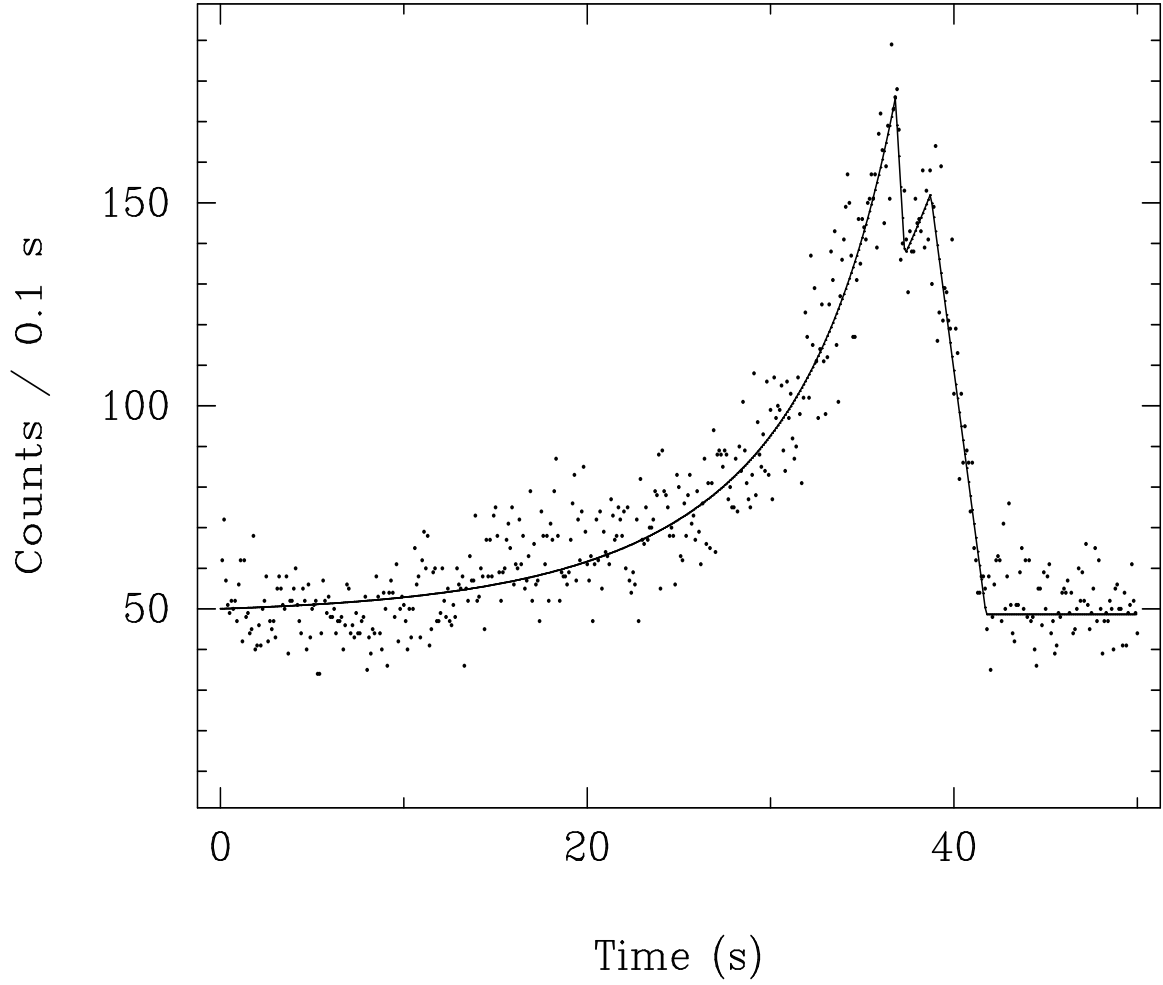} 
\caption{Profile of a regular burst is shown along with
the best fit model (continuous line) consisting of two burst with exponential
rise and linear decay, and a constant intensity.}
\end{figure}

\begin{figure}
\vskip 16cm
\includegraphics{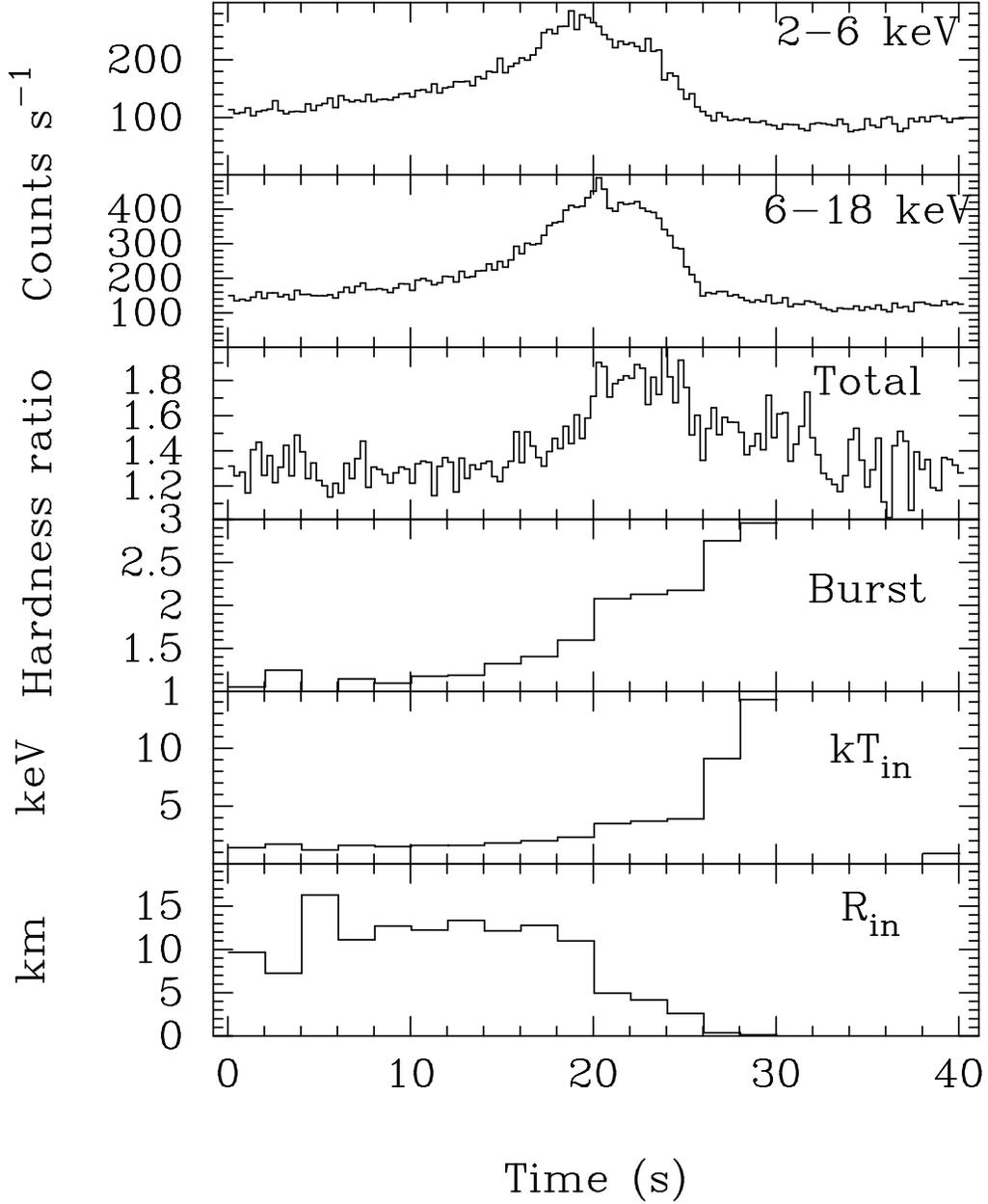} 
\caption{The burst profile in two different energy ranges
are shown in the top two panels. The middle two panels show the hardness
ratio of the total light-curve and hardness ratio of only the burst.
The quiescent intensity (30$-$40 s time range in the figure) was
subtracted from the light curves and the resultant profiles were
taken to generate the hardness ratio of only the bursting component. This
is why the fourth panel extends only upto 30 s. The temperature (kT$_{in}$)
and radius (R$_{in}$) of the innermost disk of an assumed multi-temperature
disk emission, are plotted in the bottom two panels.}
\end{figure}

\end{document}